\documentclass[twocolumn]{aastex701}

\usepackage{amsmath}
\usepackage{multirow}
\usepackage{subfig}

\newcommand\Hii{H~\textsc{ii}}
\newcommand\Nii{[N~{\sc ii}]}

\shorttitle{Low-frequency HRRLs}
\shortauthors{Salas et al.}

\graphicspath{{.}}

\begin{document}

\title{Constraints on the properties of warm ionized gas from low-frequency hydrogen radio recombination lines} 

\correspondingauthor{Pedro Salas}
\email{psalas@nrao.edu}

\author[0000-0001-8271-0572]{Pedro Salas}
\affiliation{Green Bank Observatory,
155 Observatory Road,
Green Bank, WV 24915, USA}
\email{psalas@nrao.edu}

\author[0000-0001-6527-6954]{Kimberly L. Emig}
\affiliation{National Radio Astronomy Observatory,
520 Edgemont Road,
Charlottesville, VA 22903, USA}
\email{kemig@nrao.edu}

\author[0000-0001-8061-216X]{Matteo Luisi}
\affiliation{Westminster College, USA}
\email{luisimd@westminster.edu}

\author[0000-0002-1732-5990]{D.~Anish Roshi}
\affiliation{Florida Space Institute, University of Central Florida, Orlando, Florida 32826, USA}
\affiliation{Arecibo Observatory, Arecibo, Puerto Rico 00612, USA}
\affiliation{Center for Advanced Research in Science and Engineering, University of Puerto Rico,
Mayag\"uez, PR 00681-9000, USA}
\email{}


\author[0000-0001-8800-1793]{Loren Anderson}
\affiliation{West Virginia University, USA}
\email{}

\begin{abstract}
The ionized gas in the Milky Way is a major component of the interstellar medium. Observations of extinction free tracers, such as hydrogen radio recombination lines (HRRLs), have revealed the presence of a dense (electron density $1$ to $100$~cm$^{-3}$) warm ionized medium. Motivated by advances in radio instrumentation, the existence of fully sampled HRRL maps, and a better knowledge about the population of discrete \Hii\ regions in our Galaxy, we have acquired new low-frequency ($\nu\lesssim1$~GHz) observations of HRRLs to characterize the properties of this gas. We target three positions in the Galactic plane, with few or no known \Hii\ regions, using the $342$~MHz and $800$~MHz feeds of the Green Bank Telescope. We detect HRRL emission from all three positions. We combine these with the fully sampled HRRL $5.8$~GHz cubes from the GBT Diffuse Ionized Gas Survey (GDIGS) to determine the gas properties using a forward modeling approach. From our analysis we find electron densities between $6$ and $15$~cm$^{-3}$, and that to determine the gas temperature and emission measure we require informative priors or higher signal-to-noise observations.
\end{abstract}


\section{Introduction} \label{sec:intro}

Radiative feedback from hot massive stars plays an important role in setting the characteristics of the interstellar medium (ISM). The Lyman continuum (LyC) radiation from these massive stars creates pockets of warm ionized gas surrounding them, \Hii\ regions, and on larger scales this radiation is the likely source of ionization for the warm ionized medium \citep[WIM, for a review see][]{Haffner2009}. As a direct product of stellar feedback, the properties of ionized gas offer insights about the sources of ionization and the impacts of stellar feedback on the ISM.

The existence of the WIM was first proposed by \citet{Hoyle1963} as a layer extending $\sim1$~kpc from the Galactic plane with an electron density $n_{\mathrm{e}}\sim0.1$~cm$^{-3}$ and a temperature $\sim10^{4}$~K \citep[see also][]{Reynolds1973}. Since then its properties have been characterized using a variety of observational methods. In particular through observations of the H$\alpha$ line at optical wavelengths and observations of pulsars \citep{Reynolds1991}. As a recombination process, the H$\alpha$ brightness is dependent on the emission measure (EM $=\int n_{\mathrm{e}}^{2}dl$), while pulsar observations provide an estimate of the dispersion measure (DM $=\int n_{\mathrm{e}}dl$). From the combined use of these two quantities it has been determined that the WIM has a low density ($n_{\mathrm{e}}\sim0.1$~cm$^{-3}$) and large volume filling factor ($f_{\mathrm{V}}>0.2$) near the WIM's scale height, $z\sim1$~kpc. From the ratios of optical lines the temperature of the WIM is constrained to be $6000$--$10000$~K \citep{Madsen2006}, in notable agreement with the earliest estimates. 

At low Galactic latitudes our knowledge of the properties of the WIM is less certain due to the presence of larger amounts of dust and the edge-on view of our Galaxy. Dust scatters and absorbs optical photons, complicating the interpretation of optical observations and precluding observations of highly obscured regions. In addition, the EM is dominated by denser gas, while the DM by lower density material. The combined effect is that different tracers might be probing the gas to different distances. Despite these complications it has been possible to characterize the properties of the WIM in the Galactic disk through observations of the thermal continuum \citep{Mezger1978, Murray2010}, radio recombination line (RRL) emission \citep{Shaver1976, Lockman1976, Anantharamaiah1985a,Anantharamaiah1986, Roshi2000, Heiles1996a, Alves2015}, far-infrared (FIR) fine structure line emission from N$^{+}$ \citep{Bennett1994,Goldsmith2015,Pineda2019,Langer2021}, and the dispersion of pulsar profiles \citep[][and references therein]{Berkhuijsen2006}.

Observations of the ionized gas in the Galactic plane suggest that $84\%$ of ionizing photons are emitted by O stars outside of compact \Hii\ regions, in gas characterized by densities of $n_{\mathrm{e}}\approx5\mbox{--}30$~cm$^{-3}$ \citep[e.g.,][]{Mezger1978}. Historically, this dense WIM (D-WIM) has been considered separately from the WIM found at higher Galactic latitudes. Various conclusions have been drawn regarding the primary origin of the D-WIM, including (i) envelopes of \Hii\ regions \citep{Shaver1976,Anantharamaiah1986,McKee1997,Roshi2001}, (ii) a pervasive component that could be an extension of the WIM \citep{Bennett1994,Heiles1996b,Berkhuijsen2006}, and (iii) originating from only a select few of the most luminous star-forming regions in the Galaxy \citep{Murray2010}. Distinguishing between these scenarios, and the potential connection between the D-WIM and WIM could be facilitated by characterizing the properties of the ionized gas in the Galactic plane as a function of position and velocity. In this work we revisit the use of hydrogen RRLs (HRRLs) for this purpose.

HRRLs (for a review see \citealt{Gordon2009}) are extinction free probes of ionized gas and its kinematics. In our Galaxy, low-frequency HRRLs exhibit maser-like emission, which makes them a powerful probe of the physical conditions of the gas; such as density, temperature and radiation field \citep[e.g.,][]{Roshi2001,Oonk2017,Emig2020b}. These properties of HRRLs have been used in the past to characterize the ionized gas in the Galactic plane, providing complementary information to that obtained through other tracers, such as the free-free continuum or FIR [NII] lines.

Using the Ooty Radio Telescope (ORT) \citet{Anantharamaiah1985a} observed 50 positions in the Galactic plane at $325$~MHz, with a beam size of $2\arcdeg$. They reported detections between Galactic longitudes of $357\arcdeg$ and $40\arcdeg$. By combining these observations with higher frequency HRRL observations at $1.4$~GHz they derived gas densities between $0.5$~cm$^{-3}$ and $10$~cm$^{-3}$, the lower values corresponding to positions assumed to be free of \Hii\ regions, and the higher values towards \Hii\ regions. Based on these results \citet{Anantharamaiah1985b} explained the low-frequency HRRL emission as arising from the low density envelopes of \Hii\ regions. Using the same telescope, \citet{Roshi2000} expanded the HRRL survey by targeting the lines at $327$~MHz with a $2\arcdeg\times2\arcdeg$ beam. Using a similar method as \citet{Anantharamaiah1985b}, \citet{Roshi2001} derived similar densities and sizes for the gas responsible for the HRRL emission. The latest observations with the ORT also derive similar gas properties \citep{Baddi2012}. 

At higher frequencies, the HRRL emission is dominated by spontaneous transitions. This makes them a direct probe of the gas temperature and emission measure. Moreover, the spatial resolution achieved at higher frequencies results in a sharper picture of the ionized gas. \citet{Alves2015} used data from the HI Parkes All-Sky Survey \citep[HIPASS ][]{Barnes2001} to derive fully sampled maps of HRRLs at $1.4$~GHz with a $14\farcm4$ beam. \citet{Anderson2021} used the Green Bank Telescope (GBT) to map HRRL emission by stacking together $15$ Hn$\alpha$ lines between $4.5$ and $7$ GHz with a $2\farcm65$ beam. This survey, the GBT Diffuse Ionized Gas Survey (GDIGS), mapped the inner Galactic plane between $32.3\arcdeg$ and $-5\arcdeg$ in Galactic longitude and $|b|\leq0.5\arcdeg$. \citet{Hong2022} used the Five-hundred-meter Aperture Spherical radio Telescope (FAST) to map HRRL emission using the Hn$\alpha$ lines between $1$ and $1.5$~GHz. 

Motivated by the existence of HRRL surveys at frequencies above $1$~GHz, a more complete census of \Hii\ regions in the Galaxy \citep{Anderson2014,Anderson2018}, the smaller beam width of the GBT for HRRL observations below $1$~GHz (relative to the ORT), and the large fractional bandwidth afforded by the Versatile GBT Astronomical Spectrometer \citep[VEGAS,][]{Prestage2015}, we carried out observations of three regions in the area covered by the GDIGS survey to characterize the properties of the ionized gas in the Galactic plane. Our aim was to understand what characteristics of the gas we can determine, given new observations and ancillary data.

This work is organized as follows: In Section \ref{sec:data} we describe new GBT observations and the ancillary data used in this analysis. In Section \ref{sec:results} we present the HRRL spectra, and radio continuum spectral energy distribution. We use the HRRL spectra and continuum emission to derive gas properties in Section \ref{sec:gasprops}. We end with a summary of our work in Section \ref{sec:summary}.

\section{Data}
\label{sec:data}


\subsection{GBT RRL Observations and Data Reduction}

\begin{figure*}
    \includegraphics[width=\textwidth]{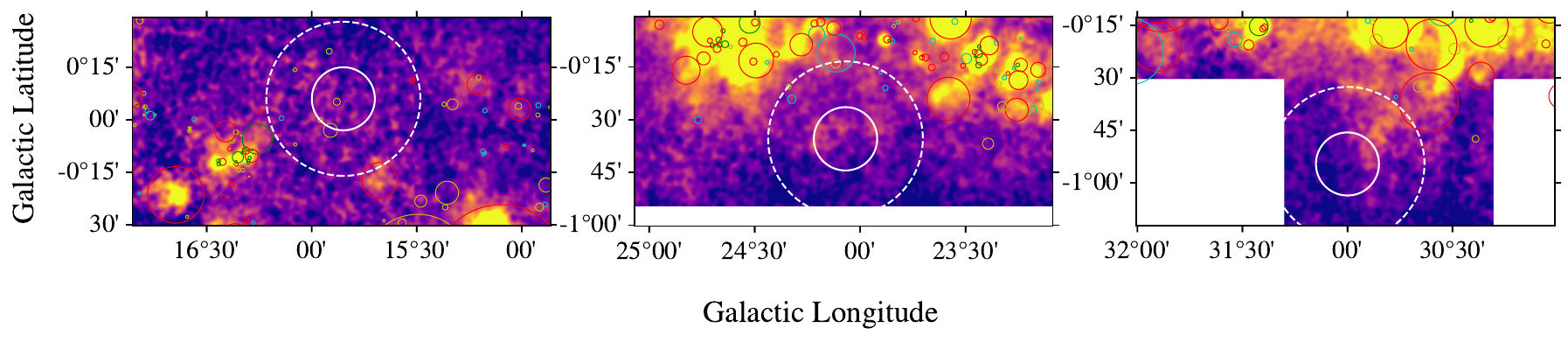}
    \caption{Regions observed with the GBT. The background images are moment 0 maps from GDIGS. The white circles show the beam size of the GBT at the lowest frequency used for the 800 MHz (solid) and 340 MHz (dashed) observations. The white circles are centered on the regions observed. The red, cyan, green and yellow circles show known, candidate, group and radio quiet \Hii\ regions, respectively, from V2.3 of the WISE catalog of \Hii\ regions \citep{Anderson2014}.}
    \label{fig:obs}
\end{figure*}

We used the GBT between January and February $2020$, to observe three regions in the Galactic plane using the $342$~MHz and $800$~MHz prime focus (PF) receivers under project AGBT19B\_334. The regions observed are presented in Table~\ref{tab:avgspec}. These regions were chosen to minimize their overlap with known \Hii\ regions in the Galactic plane \citep{Anderson2014,Anderson2018}. 
Around G15.85+0.10 there are five known \Hii\ regions within a $36\arcmin$ diameter (the size of the half power beam width of the GBT at $342$~MHz), while for the other two regions there are no known \Hii\ regions within this area. Maps of the regions are presented in Figure~\ref{fig:obs}.

We employed the total-power mode with position-switching for all observations, and VEGAS as backend. The integration time per spectral dump was set to $30$~s, the spectral windows to a bandwidth of $23.44$~MHz and $32768$ channels. At $800$~MHz we used $15$ spectral windows to cover the frequency range between $673$ and $925$~MHz. In this frequency range there are $22$ H$n\alpha$\footnote{Transitions involving a change in principal quantum\\ number $\Delta n=1$.} lines with principal quantum numbers $n=192$ to $213$. At $342$~MHz we used $7$ spectral windows to cover the frequency range between $283$ and $391$~MHz, where there are $29$ H$n\alpha$ lines ($n=256$ to $284$).

At the beginning of each session we corrected the pointing of the telescope and observed J1720--0058 (3C353) as an absolute flux calibrator. We used the flux densities measured by \citet{Perley2017} to determine the temperature of the noise diode. We assume a $10\%$ uncertainty in the temperature calibration, an aperture efficiency of $0.71$ and a main beam efficiency of $0.91$ \citep{Maddalena2010,Maddalena2012} to convert from antenna to main beam temperature.

The data reduction was carried using custom \texttt{Python} scripts based on \texttt{dysh}\footnote{We used version 0.4.0a of \texttt{dysh}.} \citep{Pound2025}. We use the notation of \citet{Winkel2012} to describe the data calibration steps.

For each spectral window and polarization we determine the $GT_{\rm{cal}}$ product for each integration from the difference between $P^{\rm{cal}}$ and $P$. Here $G$ is the gain of the telescope, $T_{\rm{cal}}$ the temperature of the noise diode, and $P$ is the recorded power. We smooth $GT_{\rm{cal}}$ along the frequency axis and model it using a linear function of frequency. Then we calibrate the data to antenna temperature units by dividing by $GT_{\rm{cal}}$ and multiplying by $T_{\rm{cal}}$.

We identify and blank radio frequency interference (RFI) in the calibrated spectra. To identify RFI we subtract a smooth function from each spectrum. The smooth function is derived using the ArPLS method \citep{Baek2015} with a smoothing factor of $10^{5}$. From the subtracted spectra we estimate the rms as a function of frequency using a rolling window. The subtracted spectra are then divided by the rms, and a SumThreshold method \citep{Offringa2010} is applied to this rms normalized data to identify RFI. Any channels flagged as RFI by this method are then blanked.

After RFI blanking we time average the data for each source using the inverse variance as weight for each spectrum. To estimate the variance we use the average of the rms computed during the RFI identification step. 

After time averaging we are left with one spectrum per spectral window and polarization for each source, receiver and observing session. Each one of these spectra is further split into sub-windows centered on the rest frequencies of the H$n\alpha$ RRLs present in the frequency ranges of the receivers. Each sub-window is $1000$~km~s$^{-1}$ wide.


All the time averaged H$n\alpha$ spectra are inspected for leftover RFI and strong deviations from a flat frequency response. The spectra that are clean from RFI and show a frequency response that can be modeled by a polynomial of order $\leq5$ are corrected by fitting a polynomial to line-free channels and then subtracting the best fit polynomial from the spectra. We investigated the effect of the polynomial fitting process by adjusting the order of the polynomial and the range of line-free channels. We find that the changes in the derived line properties varied by less than $3\%$. The baseline corrected spectra for a source are then interpolated to a common velocity grid and averaged together using the inverse of the variance computed in line free channels as weight \citep[e.g.,][]{Emig2020}. After this, we are left with a single H$n\alpha$ spectrum for each source and receiver. The properties of the averaged spectra are presented in Table~\ref{tab:avgspec}.

For the observations with the $800$~MHz observations, the difference between the observed and theoretical spectral noise in the averaged spectra is less than $30\%$. For the $342$~MHz observations this difference is less than $50\%$. In both cases the observed noise levels are higher than the theoretical values. We note that for computing the theoretical noise we did not take into account the details of how the bright continuum from the Galactic plane couples to the antenna through its sidelobes. This means that the theoretical noises are underestimates.

\begin{deluxetable*}{lcccccccc}
\tablecaption{\label{tab:avgspec} Properties of the averaged H$n\alpha$ spectra.}
\tablewidth{\textwidth}
\tablehead{
\colhead{Region} & \colhead{$\ell$} & \colhead{$b$} & \colhead{\# lines} & \colhead{$\langle n\rangle$} & \colhead{$\theta_{\rm{HPBW}}$} & \colhead{$\Delta v$}    & \colhead{$t_{\rm{ON}}$} & \colhead{noise} \\
       & ($\arcdeg$) & ($\arcdeg$) &          &     & ($\arcmin$)          & (km s$^{-1}$) & (minutes)  & (mK) 
       }
\colnumbers
\startdata
\multicolumn{9}{c}{$342$~MHz} \\
\hline
G15.85+0.10  & $15.85$ & $0.10$ & $27$ & $268$ & $36.27$ & $0.54$ & $35.2$  & $90$ \\
G24.07$-$0.59  & $24.07$ & $-0.59$ & $17$ & $269$ & $36.67$ & $0.54$ & $108.9$ & $60$  \\
G31.00$-$0.91  & $31.00$ & $-0.91$ & $11$ & $266$ & $35.46$ & $0.54$ & $152.4$ & $61$  \\
\hline
\multicolumn{9}{c}{$800$~MHz} \\
\hline
G15.85+0.10  & $15.85$ & $0.10$ & $17$ & $203$ & $15.79$ & $0.32$ & $11.7$ & $17$ \\
G24.07$-$0.59  & $24.07$ & $-0.59$ & $24$ & $203$ & $15.79$ & $0.32$ & $33.2$ & $10$ \\
G31.00$-$0.91  & $31.00$ & $-0.91$ & $20$ & $203$ & $15.79$ & $0.32$ & $23.4$ & $9$  \\
\enddata
\tablecomments{
(4) Number of lines used in the stack.\\
(5) Weighted average principal quantum number of the stacked lines.\\
}

\end{deluxetable*}

\subsection{Higher frequency HRRLs}

We complement our RRL observations with observations of HRRLs at higher frequencies. We use data from the GBT Diffuse Ionized Gas Survey \citep[GDIGS,][]{Anderson2021}\footnote{The cubes were downloaded from the GDIGS survey data server, \url{http://astro.phys.wvu.edu/gdigs/}, on April 2021.} The GDIGS HRRL cube is an average of $15$ Hn$\alpha$ transitions between $4.7$ and $7.3$ GHz. The averaged Hn$\alpha$ line has a mean frequency of $5757.8$ MHz, roughly corresponding to H$104\alpha$. The averaged cube has a spatial resolution of $2\farcm65$ and a channel spacing of $0.5$~km~s$^{-1}$. The temperature scale of the GDIGS observations was calibrated against observations of 3C286 using the \citet{Ott1994} flux scale, and it has a $5\%$ error. The differences in flux between the \citet{Ott1994} and \citet{Perley2017} scales are $\approx2\%$.

\subsection{Continuum}

    

We use the following continuum data to derive radio spectral energy distributions (SEDs) for the observed regions: Plank $70.4$~GHz map\footnote{\url{https://irsa.ipac.caltech.edu/data/Planck/release_3/all-sky-maps/}} at $13.22\arcmin$ resolution \citep{PlanckCollab2020b}, the Parkes $6$~cm ($4$~GHz) survey at a resolution of $4\farcm2$ \citep{Haynes1978}, the S-band Polarization All-Sky Survey (S-PASS) at $2.3$~GHz and $8.9\arcmin$ resolution \citep{Carretti2019}, the $1.4$~GHz Continuum processed from Parkes surveys in L-band at $14.4\arcmin$ resolution \citep{Calabretta2014}, the $408$~MHz all sky survey at $56\arcmin$ resolution \citep{Haslam1981, Haslam1982} destripped by \citet{Remazeilles2015}, and the OVRO-LWA map at $57.5$~MHz and $15.5\arcmin$ resolution \citep{Eastwood2018}. 

We extract $2\degr\times2\degr$ cutouts around the areas observed in RRLs (Table~\ref{tab:avgspec}) and grid the images to a pixel size of $30\arcsec$. Then, we convolve these cutouts to a resolution of $16\arcmin$, except for the Haslam $408$~MHz data which has a $56\arcmin$ resolution.

\section{Results}
\label{sec:results}

\subsection{RRL emission}
\label{ssec:rrlemi}

\begin{figure*}[h]
\resizebox{\hsize}{!}
{\includegraphics{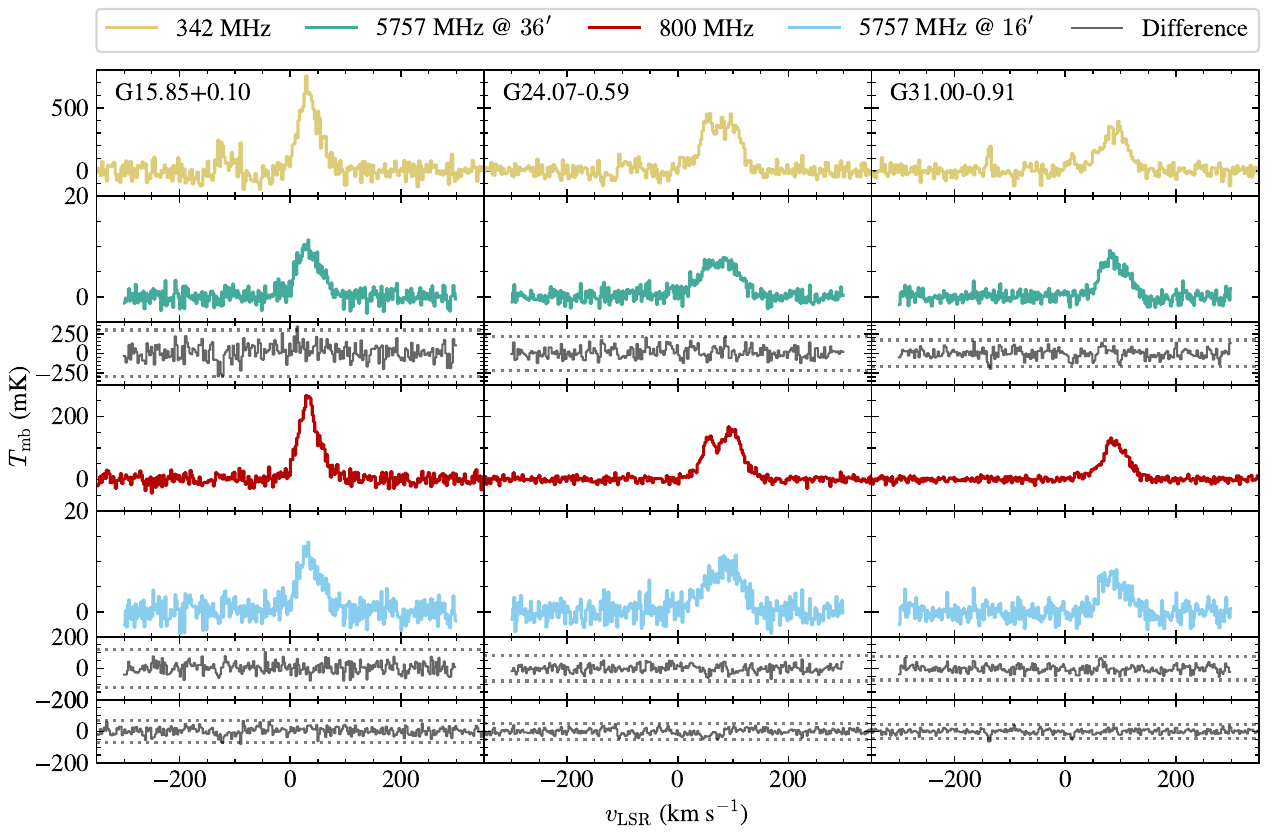}}
\caption{RRL spectra towards G15.85+0.10 (left), G24.07$-$0.59 (center) and G31.00$-$0.91 (right). From top to bottom each rows shows: the PF $342$~MHz RRL spectra at $36\arcmin$ resolution, the GDIGS spectra at $36\arcmin$ resolution, the difference between the PF $342$~MHz and GDIGS spectra at $36\arcmin$ resolution, the PF $800$~MHz RRL spectra at $16\arcmin$ resolution, the GDIGS spectra at $16\arcmin$ resolution, the difference between the PF $800$~MHz and GDIGS spectra at $16\arcmin$ resolution, and the difference between the PF $342$~MHz and $800$~MHz RRL spectra, at different resolutions.
The dotted lines show the $3\sigma$ rms of the differences. To compute the differences between the HRRL spectra at different frequencies, the higher frequency data have been scaled to match the peak of the lower frequency data.}
\label{fig:spec}
\end{figure*}

Hydrogen radio recombination lines (HRRLs) are detected from the three positions observed (Table~\ref{tab:avgspec}). The RRL spectra are presented in Figure~\ref{fig:spec}, where we also show the HRRLs from the GDIGS cubes towards the same positions with a spatial resolution matched to that of the $342$~MHz and $800$~MHz observations. The $342$~MHz spectra show the presence of helium and/or carbon RRLs, offset by $-122$~km~s$^{-1}$ and $-149$~km~s$^{-1}$ from the HRRLs, respectively.

The velocities of the HRRL emission agree across frequencies for a given position. Towards G15.85+0.10 and G31.00$-$0.91 the spectra can be described by a single velocity component, while towards G24.07$-$0.59 the $800$~MHz HRRL shows a double peaked profile, indicative of the presence of, at least, two velocity components along the line of sight. The full widths at half maximum (FWHM) of the velocity components are between $30$ and $50$~km~s$^{-1}$, based on Gaussian fitting of the line profiles (Table~\ref{tab:gbfit}). By comparison the median of the HRRL FWHM from GDIGS is $30.6$~km~s$^{-1}$ \citep{Anderson2021}. Broad line widths are expected if gas motions contribute significantly to the line width.

In Figure~\ref{fig:spec} we also show the difference between the HRRLs at different frequencies. The goal is to compare the shapes of the line profiles, ignoring differences in their relative brightness. Since the brightness of the HRRLs changes with frequency, we scale the higher frequency HRRLs to match the peak brightness of the lower frequency HRRLs when computing the differences. To compare on the same velocity axis, we average the spectra to the same velocity resolution and interpolate to a common velocity grid. For all positions, the rms of the difference over line-free channels ($-350$ to $-200$~km~s$^{-1}$ and $200$ to $350$~km~s$^{-1}$) is consistent with what is expected from the rms in the individual spectra. We also compare the rms in the difference with the properties of the channels where there is HRRL emission ($0$ to $150$~km~s$^{-1}$). Overall, the differences are mostly consistent with noise, but there is at least one channel in each region with a signal-to-noise ratio (S/N) of $\approx3$ (e.g., in the difference between the $800$~MHz and GDIGS spectra for G31.00$-$0.91, right column and sixth row in Figure~\ref{fig:spec}, around $60$~km~s$^{-1}$). Thus, although the HRRL spectra are similar, their shapes are not identical.

\begin{figure*}[h]
  \resizebox{\hsize}{!}
  {\includegraphics{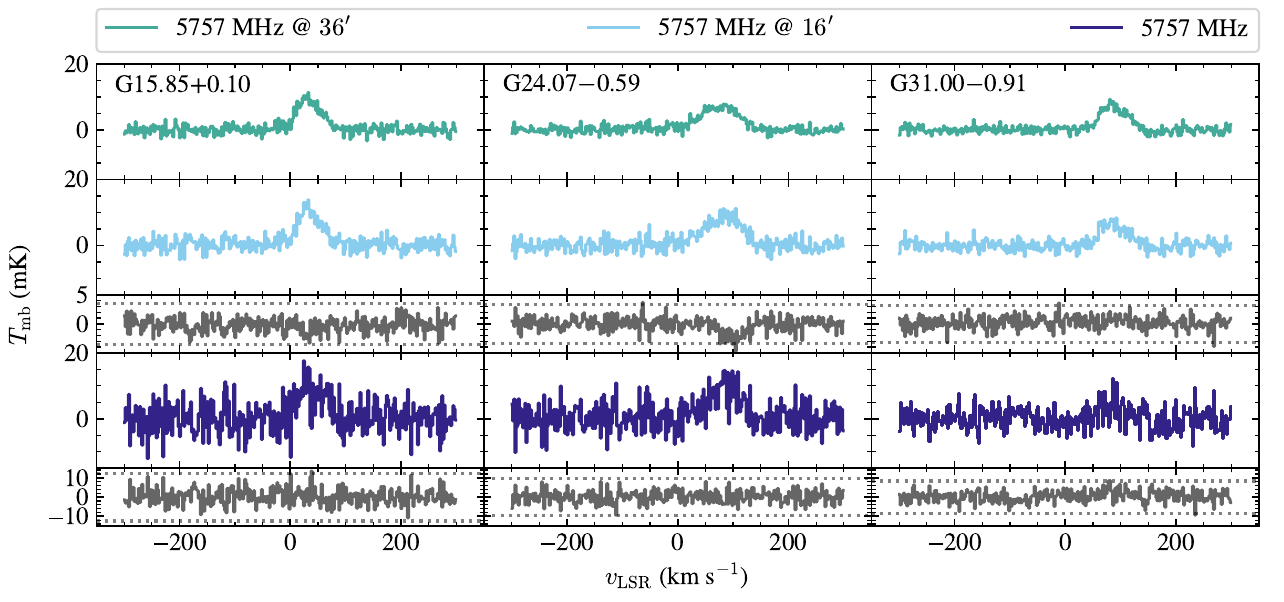}}
  \caption{GDIGS RRL spectra towards G15.85+0.10 (left), G24.07$-$0.59 (center) and G31.00$-$0.91 (right) using three different beam sizes, $36\arcmin$, $16\arcmin$ and $2\farcm65$. From top to bottom each row shows: the GDIGS spectra at $36\arcmin$ resolution, the GDIGS spectra at $16\arcmin$ resolution, the difference between the GDIGS spectra at $36\arcmin$ and $16\arcmin$ resolutions, the GDIGS spectra at $2\farcm65$ resolution, and the difference between the GDIGS spectra at $16\arcmin$ and $2\farcm65$ resolutions. The dotted lines show the $3\sigma$ rms of the differences.
  No scaling has been applied to compute the differences.}
  \label{fig:gdigsspecs}
\end{figure*}

We also use the GDIGS data to compare the HRRLs at different spatial resolutions. As the effective $\theta_{\rm{HPBW}}$ of the observations changes by a factor of $\approx2.25$ between $342$~MHz and $800$~MHz, the gas volumes probed are different. In Figure~\ref{fig:gdigsspecs} we compare the GDIGS HRRL spectra at the spatial resolutions of the $342$~MHz and $800$~MHz observations (Table~\ref{tab:avgspec}), and at the native resolution of the GDIGS observations ($2\farcm65$). To generate the GDIGS spectra at the spatial resolution of the $342$~MHz and $800$~MHz observations, we first convolve the GDIGS cube to the corresponding spatial resolution and then extract the spectra from an aperture equal to $\theta_{\rm{HPBW}}$. For G24.07$-$0.59 and G31.00$-$0.91 this misses $\approx3\%$ and $\approx5\%$, respectively, of the solid angle of the $342$~MHz beam since the aperture extends beyond the coverage of the GDIGS cubes (see Figure~\ref{fig:obs}). The difference between the HRRLs from G15.85+0.10 and G31.00$-$0.91 are consistent with noise. This indicates that there are no additional velocity components over a larger aperture, and that there is no significant beam dilution. Based on this result, it seems reasonable to consider that in G15.85+0.10 and G31.00$-$0.91 the gas properties are similar at the two spatial resolutions being considered. To simplify the analysis for these two regions, we use the average between the two GDIGS spectra extracted from the $16\arcmin$ and $36\arcmin$ apertures.
Towards G24.07$-$0.59, the high velocity HRRL (component ID 2 in Table~\ref{tab:gbfit}) is brighter at $2\farcm65$ resolution than at $36\farcm67$. The GDIGS data, at its native resolution, shows that at $\approx90$~km~s$^{-1}$ the emission is brighter over a $\approx5\arcmin\times5\arcmin$ region towards $(\ell,b)=24.18\arcdeg,-0.53\arcdeg$. The emisison from this are dominates the emission of the $90$~km~s$^{-1}$ velocity component towards G24.07$-$0.59, which is beam diluted at the resolution of the $342$~MHz observations.

\begin{figure*}[h]
  \resizebox{\hsize}{!}
  {\includegraphics{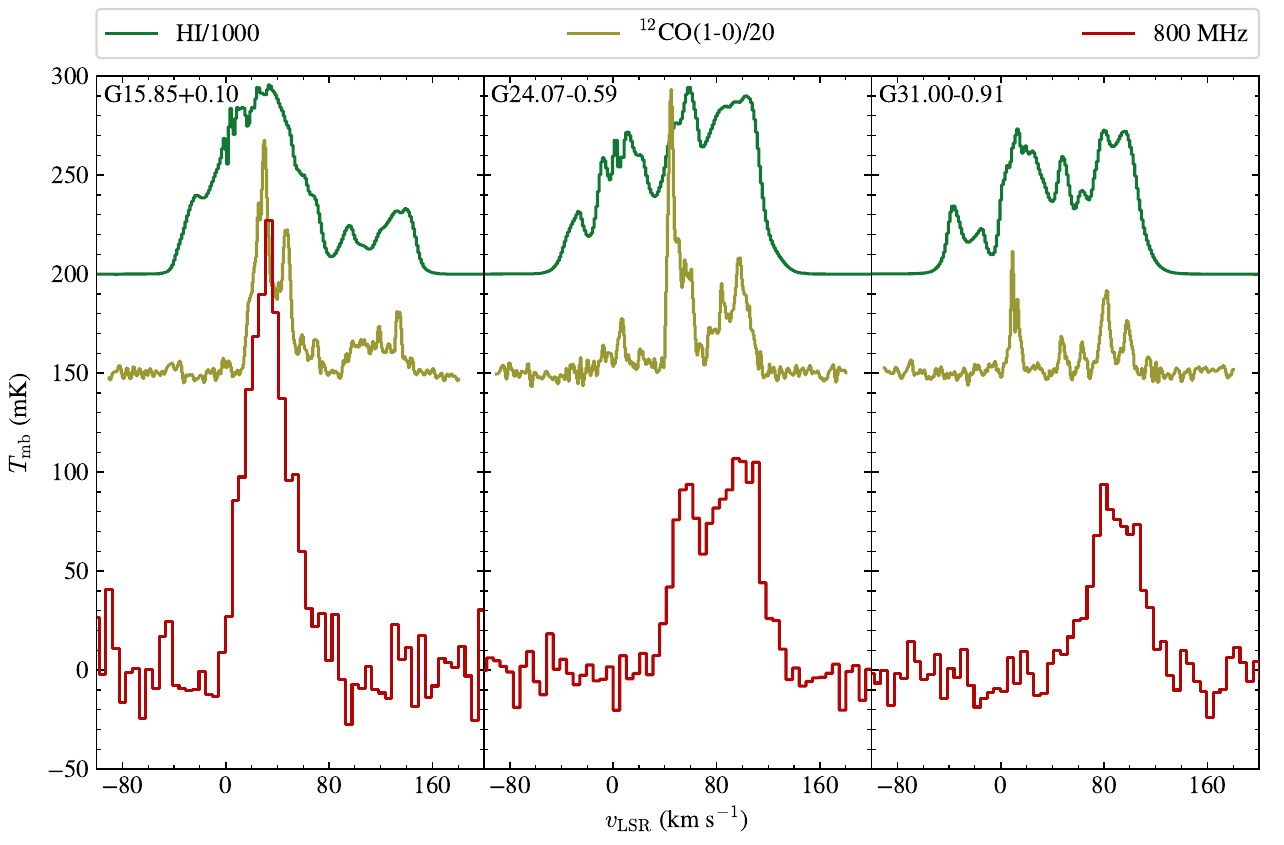}}
  \caption{Comparison between the HRRL emission at $800$~MHz, $21$~cm-HI and $^{12}$CO$(1\mbox{--}0)$. The HI and CO spectra were extracted from a $16\arcmin$ aperture, to match the resolution of the HRRL observations at $800$~MHz.}
  \label{fig:gases}
\end{figure*}

We compare the HRRL profiles at $800$~MHz with $21$~cm-HI and $^{12}$CO$(1\mbox{--}0)$ line emission in Figure~\ref{fig:gases}. The HRRL emission is confined to velocity ranges where we also find CO emission, but there is CO and HI emission over a wider range of velocities. The correspondence between the HRRL and CO emission suggests that the gas being probed by the HRRLs is associated with the spiral arms of the Galaxy. The CO profiles for all three positions also show the presence of multiple velocity components within the velocity ranges where HRRL emission is detected. Thus, it is possible that the HRRL profiles are a blend of multiple velocity components along the line of sight. We also note that there is no correlation between the brightness of the CO line and the HRRL at $800$~MHz.

To measure the line properties we fit the spectra using Gaussian line profiles. To determine the number of velocity components to use during the fitting process we inspect the spectra visually for the presence of multiple peaks and deviations from a line profile described by a single Gaussian. During the visual inspection we also define the initial values for the parameters used during the fit. We use the $800$~MHz HRRLs to inspect the profiles and determine how many velocity components to include in the line fit at the other frequencies , since these are the spectra with the highest S/N. Although, the number of velocity components need not be the same at different frequencies, using the same number of velocity components at different frequencies simplifies our analysis, specially since the method we use to derive gas properties (Section~\ref{ssec:method}) does not consider the gas dynamics, nor multiple density layers. For G15.85+0.10 a single velocity component provides the lowest Akaike information criterion \citep[AIC,][]{Akaike1974}, for G24.07$-$0.59 two components, and for G31.00$-$0.91 a single component. Thus, we use a single velocity component when fitting the profiles for G15.85+0.10 and G31.00$-$0.91, and two for G24.07$-$0.59. For G24.07$-$0.59 fitting the GDIGS and the 342~MHz spectra using two velocity components is degenerate and results in large uncertainties for the best fit parameters given the S/N of the spectra. The best fit parameters are presented in Table~\ref{tab:gbfit}.

\begin{deluxetable}{lccccc}
\digitalasset
\tablecaption{\label{tab:gbfit} Best fit HRRL properties.}
\tablewidth{\textwidth}
\tablehead{
Frequency & ID & $T_{\mathrm{mb}}$ & $v_{\mathrm{c}}$ & $\Delta v$    & S/N \\
(MHz)     &    & (mK)              & (km s$^{-1}$)    & (km s$^{-1}$) &     
       }
\colnumbers
\startdata
\multicolumn{6}{c}{G15.85+0.10} \\
\hline
5757                     & 1 & $8.5\pm1.4$  & $39\pm4$     & $54\pm9$     & $6.8$\\
5757$^{\dagger}$         & 1 & $10.3\pm0.9$ & $34\pm1$     & $48\pm3$     & $19$ \\
5757$^{\ddagger}$        & 1 & $9.2\pm0.7$  & $33\pm1$ & $49\pm3$     & $24$ \\
5757$^{\dagger\ddagger}$ & 1 & $9.5\pm0.7$  & $33\pm1$     & $49\pm3$     & $24$ \\
800                      & 1 & $240\pm4$    & $32\pm1$     & $43\pm1$     & $63$ \\
342                      & 1 & $611\pm15$   & $33\pm1$     & $44\pm1$     & $39$ \\
\hline
\multicolumn{6}{c}{G24.07$-$0.59} \\
\hline
5757              & 1 & $3.0\pm1.3$  & $36\pm19$    & $45\pm35$    & -- \\
5757              & 2 & $11\pm1$     & $89\pm6$     & $49\pm11$    & -- \\ 
5757$^{\dagger}$  & 1 & $4.2\pm1.6$  & $49\pm3$     & $17\pm2$     & -- \\
5757$^{\dagger}$  & 2 & $9.0\pm0.8$  & $93\pm6$     & $22\pm4$     & -- \\
5757$^{\ddagger}$ & 1 & $4.5\pm1.0$  & $49\pm3$     & $17\pm7$     & -- \\
5757$^{\ddagger}$ & 2 & $6.5\pm0.5$  & $90\pm5$     & $22\pm3$     & -- \\
800               & 1 & $114\pm6$    & $54\pm1$     & $29\pm3$     & -- \\
800               & 2 & $155\pm5$    & $97\pm1$     & $45\pm3$     & -- \\
342               & 1 & $380\pm14$   & $55\pm1$     & $36\pm4$     & -- \\
342               & 2 & $377\pm12$   & $97\pm4$     & $39\pm6$     & -- \\
\hline
\multicolumn{6}{c}{G31.00$-$0.91} \\
\hline
5757                     & 1 & $4\pm1$     & $89\pm7$     & $56\pm16$    & $4$  \\
5757$^{\dagger}$         & 1 & $6.7\pm0.6$ & $87\pm2$     & $54\pm4$     & $16$ \\
5757$^{\ddagger}$        & 1 & $7.3\pm0.5$ & $87\pm1$     & $54\pm3$     & $27$ \\
5757$^{\dagger\ddagger}$ & 1 & $7.1\pm0.5$ & $87\pm1$     & $54\pm3$     & $25$ \\
800                      & 1 & $116\pm8$   & $89\pm1$     & $53\pm3$     & $70$ \\
342                      & 1 & $321\pm9$   & $88\pm2$     & $56\pm4$     & $36$ \\
\enddata
\tablecomments{
(2) Velocity component identifier.\\
(5) Full width at half maximum of a Gaussian line profile.\\
(6) Signal-to-noise ratio for the line intensity \citep[$f_{L}$ in ][]{Lenz1992}. For G24.07$-$0.59 we do not provide S/N values as this is a double peaked profile.\\
$^{\dagger}$ Extracted from a $16\arcmin$ beam.\\
$^{\ddagger}$ Extracted from a $36\arcmin$ beam.\\
$^{\dagger\ddagger}$ Average of the spectra extracted from the $16\arcmin$ and $36\arcmin$ beams.
}
\end{deluxetable}

For each target, we compare the best fit line properties at equivalent angular resolution. The best fit line centroids are consistent between the different frequencies, with differences of less than $3\sigma$ in all cases. This suggests that the different transitions are probing similar gas volumes or that there are no large velocity gradients between the gas at different densities, as non-LTE effects preferentially amplify HRRL emission from lower density gas at lower frequencies \citep[e.g.,][]{Shaver1976}. For G15.85+0.10 and G31.00$-$0.91, the line widths are consistent between the different frequencies, so there is no significant pressure broadening, by collisions or the radiation field. This is similar to the findings of \citet{Pedlar1978}, who reported no significant pressure broadening towards two \Hii\ regions and the Galactic Center between $242$ and $408$~MHz. The line brightness increases towards the lower frequencies. In LTE, and assuming that pressure broadening is not significant, the LTE brightness ratios H$203\alpha$/H$104\alpha$ and H$268\alpha$/H$104\alpha$ are $7.38$ and $16.93$, respectively. The observed line ratios are ${>}16$ and ${>}45$ for the H$203\alpha$/H$104\alpha$ and H$268\alpha$/H$104\alpha$ brightness ratios, respectively, suggesting that stimulated emission contributes significantly to the line brightness.

For G24.07$-$0.59 the comparison of the best fit line properties at different frequencies is not straightforward, as we must also consider the degeneracies between the best fit parameters for the blended line profiles. The best-fit results in Table~\ref{tab:gbfit} suggest that the line at $\approx50$~km~s$^{-1}$ (component $1$) does not change its width, but the second component at $\approx90$~km~s$^{-1}$ (component $2$) gets broader at lower frequencies.

In summary, HRRLs are detected towards the three positions observed at $342$~MHz, $800$~MHz, and $5757$~MHz. For G15.85+0.10 and G31.00$-$0.91, the line profiles show consistent velocity centroids and line widths, despite the changes in angular resolution. For G24.07$-$0.59 we find that the HRRL spectra are best described by two velocity components, with a brightness and width that depends on the size of the aperture. In Section~\ref{sec:gasprops} we will model the emission from these regions as a single volume.

\subsection{Distances}
\label{ssec:dist}

We estimate the distances to the gas responsible for observed HRRL emission using the line kinematics (Table~\ref{tab:gbfit}). To estimate distances from the line kinematics we use the kinematic distance (KD) calculator of \citet{Wenger2018}, method C with the \citet{Reid2014} Galactic rotation curve. To resolve the kinematic distance ambiguity we search the WISE catalog of \Hii\ regions \citep{Anderson2014} for regions with known distances which have projected distances of less than $2\arcdeg$ to our targets, and whose radial velocity is within $\pm10$~km~s$^{-1}$ from that of the HRRLs towards our targets. We also note that since the HRRL emission is stimulated, the gas is more likely to lie at the near KD, as at near distance there is more background continuum.

The HRRLs towards G15.85+0.10 have a radial velocity of $\approx33$~km~s$^{-1}$. This corresponds to a Galactocentric distance of $5.49^{+0.24}_{-0.32}$~kpc. For this position and velocity, the near and far KDs are $3.07^{+0.27}_{-0.27}$~kpc and $12.93^{+0.39}_{-0.36}$~kpc, respectively, and the tangent point velocity is $119.8^{+8.2}_{-16.3}$~km~s$^{-1}$. The nearby \Hii\ regions M16, M17 and Sh2-48 have velocity measurements and distance estimates. The \Hii\ region Sh2-48 (G016.648-00.357) shows HRRL emission at a similar radial velocity and is estimated to lie at a distance of $3.59$~kpc, based on Gaia Early Data Release 3 (EDR3) parallaxes \citep{Mendez-Delgado2022}. The same method puts M16 and M17 at distances of $1.71$~kpc and $1.82$~kpc, respectively. Based on the similar radial velocity and projected proximity between Sh2-48 and G15.85+0.10, we put G15.85+0.10 at the near kinematic distance.

The HRRLs towards G31.00$-$0.91 have a radial velocity of $89$~km~s$^{-1}$, putting the gas at a Galactocentric distance of $4.65^{+0.24}_{-0.22}$~kpc. The near and far KD are $5.02^{+0.77}_{-0.35}$~kpc and $9.03^{+0.61}_{-0.55}$~kpc, respectively. The velocity of the tangent point in this direction is $104.8^{+8.3}_{11.0}$~km~s$^{-1}$. G31.00$-$0.91 is close to W43, which shows HRRL emission at a velocity of $\approx100$~km~s$^{-1}$ \citep{Luisi2020}. Parallax measurements put W43 at a distance of $5.49$~kpc \citep{Zhang2014}. The projected and kinematic proximity of G31.00$-$0.91 and W43 suggests that the ionized gas in both regions is associated, thus we place G31.00$-$0.91 at the near KD.

For G24.07$-$0.59 we have velocities of $54$ and $97$~km~s$^{-1}$. For these velocities the Galactocentric distances are $5.27^{+0.28}_{-0.25}$~kpc and $3.86^{+0.26}_{-0.26}$~kpc, respectively. For the $54$~km~s$^{-1}$ component we have near and far kinematic distances of $3.51^{+0.36}_{-0.26}$~kpc and $11.66^{+0.40}_{-0.42}$~kpc, respectively. For the $97$~km~s$^{-1}$ component we have distances of $5.60^{+0.62}_{-0.41}$~kpc and $9.48^{+0.57}_{-0.54}$~kpc. For this position we do not attempt to resolve the kinematic distance ambiguity and instead we just quote the two possible solutions for each velocity component, although, as mentioned in the first paragraph of this subsection, the gas is more likely to lie at the near KD.

\begin{deluxetable}{lcccr}
\tablecaption{\label{tab:kdist} Kinematic distances.}
\tablewidth{\textwidth}
\tablehead{
Region & ID & $R_{\rm{gal}}$ & $D$   & Note \\
       &    & (kpc)          & (kpc) & }
\startdata
\hline
G15.85+0.10  & 1 & $5.49^{+0.24}_{-0.32}$ & $3.07^{+0.27}_{-0.27}$ & Near\\
\multirow{2}{*}{G24.07$-$0.59} & 1 & $5.27^{+0.28}_{-0.25}$ & & \\
             & 2 & $3.86^{+0.26}_{-0.26}$ & \\
G31.00$-$0.91 & 1 & $4.65^{+0.24}_{-0.22}$ & $5.02^{+0.77}_{-0.35}$  & Near \\
\hline
\enddata
\end{deluxetable}

\subsection{Continuum emission}
\label{ssec:contsed}

\begin{figure}[h]
  \resizebox{\hsize}{!}
  {\includegraphics{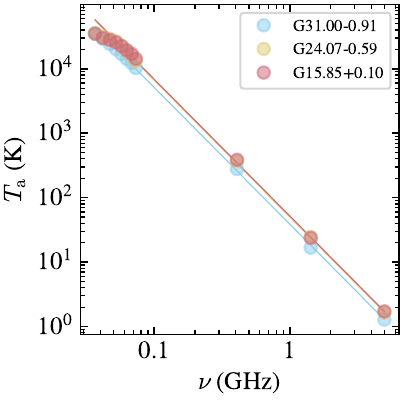}} %
  \caption{Radio continuum spectral energy distributions for the three regions observed with the GBT in RRLs. The radio continuum data has been convolved to a resolution of $56\arcmin$. The solid lines show a power-law spectra with a spectral index of $\beta=-2.2$ normalized to the brightness of the $5$~GHz data point. The data for G24.07$-$0.59 lies behind that for G15.85+0.10.}
  \label{fig:contsed}
\end{figure}

The radio continuum SEDs are presented in Figure~\ref{fig:contsed} and Table~\ref{tab:contSED}. We extract the continuum brightness towards each source within an aperture diameter of $16\arcmin$ at each frequency except for the $408$~MHz map, and additionally we extract the continuum brightness within an aperture diameter of $56\arcmin$ at all frequencies. We fit a single power-law to the brightness temperatures extracted at $16\arcmin$ and $56\arcmin$. A single power-law describes the data well without curvature in the spectrum, and the power-law slopes are consistent among the two apertures. We note that the consistency in power-law values holds whether the $57.5$~MHz (which may be missing large scale emission) and the $408$~MHz (only included in the lower resolution measurement) data points are included in the fit. 

The radio continua from these three regions is very similar, with a brightness temperature ${\approx}30$~K at $1$~GHz and ${\approx}10^{4}$~K at $58$~MHz. The high brightness of the radio continuum at low frequencies suggests that there is a significant non-thermal component to the continuum emission.

\begin{deluxetable}{lCC}
\tablecaption{\label{tab:contSED} Continuum SEDs.}
\tablewidth{\textwidth}
\tablehead{
Region &  T_0 & \beta  \\
       & (\rm{K}) &  }
\startdata
\hline
G15.85+0.10   & 32.9 \pm 1.9 & -2.20 \pm 0.01\\
G24.07$-$0.59 & 30.7 \pm 1.8 & -2.17 \pm 0.01 \\
G31.00$-$0.91  & 25.6 \pm 1.8  & -2.22 \pm 0.02  \\
\hline
\enddata
\tablecomments{ $ T(\nu)=T_0(\nu / \mathrm{GHz})^{\beta}$
}
\end{deluxetable}

\section{Ionized gas properties}
\label{sec:gasprops}

In this section we use the HRRL spectra to determine the properties of the gas, particularly its electron density. We focus on two of the three regions (G15.85+0.10 and G31.00$-$0.91) as the HRRL spectra are consistent with probing a single volume (see Section~\ref{ssec:rrlemi}). 

\subsection{Method}
\label{ssec:method}

The brightness of a recombination line is given by \citep{Shaver1975},
\begin{multline}\label{eq:TB}
    T_{\rm{b}} = T_{\rm{bg}} {\rm e}^{-\tau_{c}} \left( {\rm e}^{-\tau_{l}} - 1 \right)
    + \\ 
    T_{\rm{e}} \left[ \frac{b_n \tau_{l}^* + \tau_{c}}{\tau_{l} + \tau_{c}} \left( 1 - {\rm e}^{- \left( \tau_{l} + \tau_{c} \right)} \right) - \left( 1 - {\rm e}^{- \tau_{c}}\right) \right] \\
    + T_{\rm{nth}} \left[ \frac{1 - {\rm e}^{- \left( \tau_{l} + \tau_{c} \right) }}{\tau_{l} + \tau_{c}}  - \frac{1 - {\rm e}^{- \tau_{c}}}{\tau_{c}} \right],
\end{multline}
\noindent
where $T_{\rm{bg}}$ is the brightness of the background continuum, $T_{\rm{e}}$ the electron temperature of the gas emitting HRRLs, $T_{\rm{nth}}$ is the contribution to the continuum brightness due to non-thermal emission in the HRRL emitting gas, $\tau_{c}$ and $\tau_{l}$ are the continuum and line optical depths, respectively, $\tau_{l}^*$ is the LTE line optical depth, and $b_n$ is the ratio of the number of hydrogen atoms in level $n$ between the LTE and the non-LTE cases. In Equation~(\ref{eq:TB}) all quantities, except for the electron temperature, depend on the frequency. To compute the departure coefficients, $b_{n}$, we follow \citet{Salgado2017a}. We compute $b_{n}$ for electron densities between $0.1$~cm$^{-3}$ and $1500$~cm$^{-3}$, and electron temperatures between $1000$~K and $21000$~K.

To compute the line optical depth we use the following expression
\begin{equation}
\tau_{l}=\frac{\pi h^{3}e^{2}}{(2\pi m_{\mathrm{e}}k_{\mathrm{B}})^{3/2}m_{\mathrm{e}}c}n^{2}f_{nn^{\prime}}\frac{n_{\mathrm{e}}^{2}L}{T_{\mathrm{e}}^{3/2}}{\rm e}^{\chi_{n}}(1-{\rm e}^{-\frac{h\nu}{k_{\mathrm{B}}T_{\mathrm{e}}}}).
\end{equation}
where $L$ is the path length of the gas, $n_{\mathrm{e}}$ its electron density, $n$ the principal quantum number of the transition, $f_{nn^{\prime}}$ is the oscillator strength \citep{Menzel1968} for a transition between levels $n$ and $n^{\prime}$, and $\chi_{n}$ is the ionization potential from level $n$,  equal to $hcRZ^{2}/n^{2}k_{\mathrm{B}}T_{\mathrm{e}}$, with $R$ the Rydberg constant ($R=R_{\infty}/(1+m_{\mathrm{e}}/M_{x})$, with $M_{x}$ the nuclear mass).

The line profile of a HRRL is described by a Voigt line profile. The width of its Gaussian core is determined by the thermal and non-thermal motion of the gas, $v_{\mathrm{rms}}$. The width of the Lorentzian wings of the profiles are determined by the effects of pressure broadening, by collisions and radiation. We use the expressions of \citet{Salgado2017b} to describe the effects of pressure broadening. To evaluate the collisional broadening term we use the electron density and temperature of the gas. For radiation broadening we add the contributions from the background and non-thermal continuum and adopt a filling factor of unity. 

The line brightness, Eq.~\ref{eq:TB}, depends on $T_{\rm{bg},\nu}$ and $T_{\rm{nth}}$. The continuum SEDs only provide upper limits to them, as these provide an integrated value along the line of sight. In general, we expect that $T_{\rm{bg},\nu}=f_{\rm{bg}}T_{\rm{c}}$, with $0\leq f_{\rm{bg}}<1$, and $T_{\rm{nth}}=f_{\rm{m}}T_{\rm{c}}$, with $0\leq f_{\rm{m}}<1$, and $f_{\rm{bg}}+f_{\rm{m}}\leq1$. We adopt the prescription by \citet{Roshi2001}, setting $T_{\rm{bg},\nu}=T_{\rm{c}}(D_{\rm{gal}}-D)/D_{\rm{gal}}$ and $T_{\rm{nth}}=T_{\rm{c}}L/D_{\rm{gal}}$, with $T_{\rm{c}}$ the observed continuum brightness, $D_{\rm{gal}}$ the size of the Galactic disk along the line of sight, $D$ the distance to the gas responsible for the HRRL emission and $L$ the size of the HRRL emitting region along the line of sight. This is equivalent to assuming that the ISM is homogeneous behind the HRRL emitting region. We adopt a radius for the Galactic disk of $20$~kpc.

Once the contributions from the background and non-thermal continuum have been fixed, the model that predicts the HRRL spectra has five parameters; $v_{\mathrm{c}}$, $v_{\mathrm{rms}}$, $n_{\rm{e}}$, $T_{\rm{e}}$ and the line of sight (LoS) size of the region emitting HRRLs, $L$. Alternatively, the problem can be parametrized in terms of the emission measure: EM$\,=n_{\mathrm{e}}^{2}L$, for an homogeneous medium. The central velocity of the lines, $v_{\mathrm{c}}$, is determined from the spectra (e.g., Table~\ref{tab:gbfit}).

To derive the gas properties, we compute HRRL profiles at each principal quantum number. The peak brightness of the HRRL is set by Eq.~\ref{eq:TB}, and the width of the profile by the combination of Doppler and pressure broadening (which produces a Voigt line profile). These predicted HRRL profiles are then compared to the observed ones to determine the likelihood of the observations given the free parameters $v_{\mathrm{rms}}$, $n_{\rm{e}}$, $T_{\rm{e}}$, EM, $f_{\rm{bg}}$ and $f_{\rm{m}}$, and the observed continuum and $v_{\mathrm{c}}$, which we fix. This comparison is carried out using a Markov Chain Monte Carlo method to derive posterior distributions for the free parameters. This approach is different from that by \citet{Anantharamaiah1985b} in that we compare the line profiles instead of analyzing the line widths and line intensities separately.

\begin{figure*}[ht]
\resizebox{\hsize}{!}
{\includegraphics{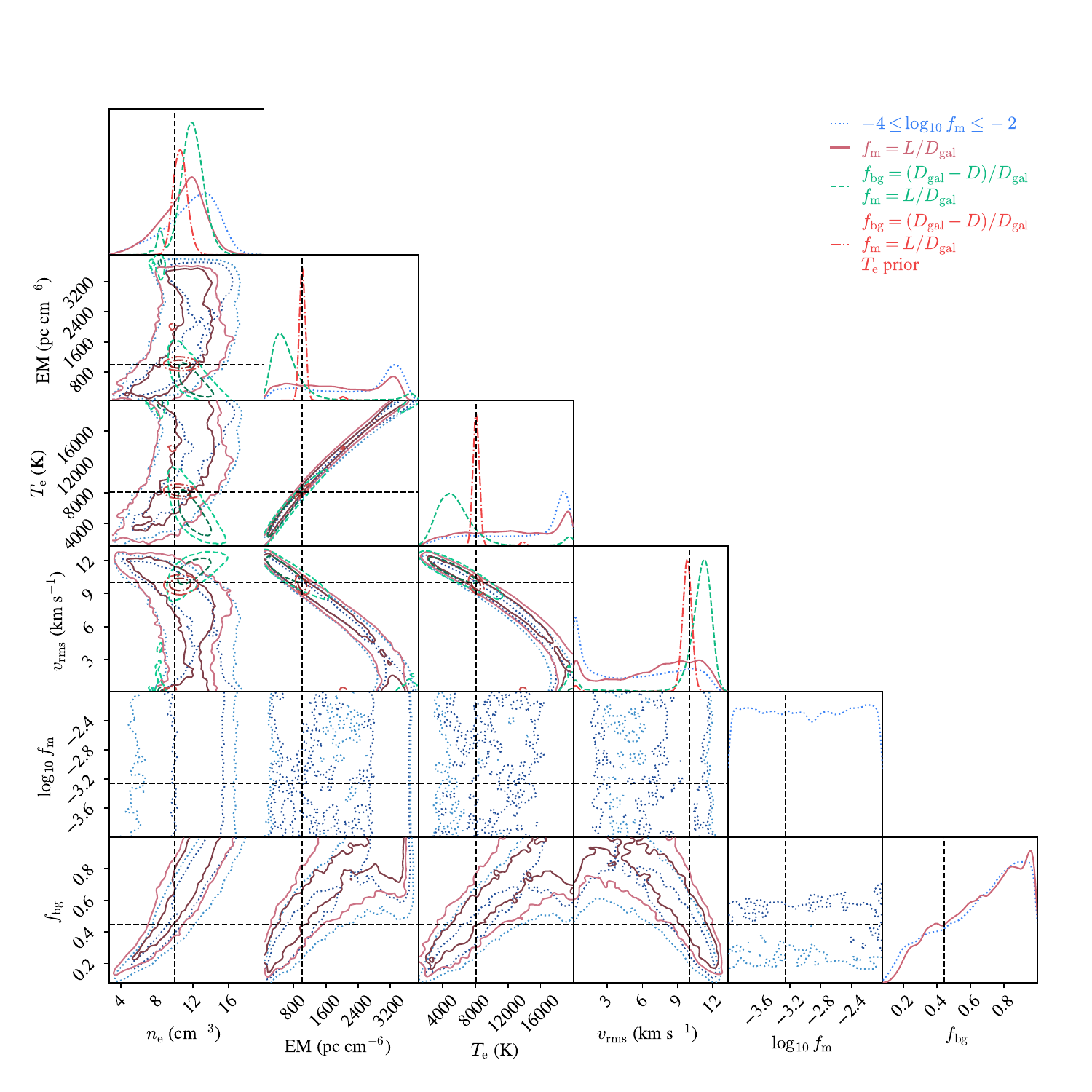}}
\caption{Posterior probability distributions for the model parameters from the validation exercise. The vertical and horizontal lines show the input values used to generate the synthetic spectra; $n_{\mathrm{e}}=10$~cm$^{-3}$, $T_{\mathrm{e}}=8000$~K, EM$\,=1000$~pc~cm$^{-6}$, $f_{\mathrm{bg}}=0.44$, and $f_{\mathrm{m}}=5.5\times10^{-4}$. $f_{\mathrm{m}}$ represents the fraction of the observed continuum that is non-thermal and comes from the same volume as the HRRLs. The blue dotted line shows the case where all model parameters are free, and $f_{\mathrm{m}}$ can take values between $10^{-4}$ and $10^{-2}$, the dark red solid line shows the case where $f_{\mathrm{m}}=L/D_{\mathrm{gal}}$, the green dashed line shows the case where $f_{\mathrm{m}}=L/D$ and $f_{\mathrm{bg}}=(D_{\mathrm{gal}}-D)/D_{\mathrm{gal}}$, and the red dot-dashed line shows the case where $f_{\mathrm{m}}=L/D$, $f_{\mathrm{bg}}=(D_{\mathrm{gal}}-D)/D_{\mathrm{gal}}$ and a Gaussian prior for the electron temperature is adopted.}
\label{fig:synth_corner}
\end{figure*}

\subsection{Method validation}
\label{ssec:validation}

Before applying the method to our observations, we validate it against synthetic observations. This also helps us understand the limitations of the method, including correlations between model parameters. We set up synthetic HRRL spectra for principal quantum numbers of $106$, $203$ and $268$. We adopt a velocity centroid of $0$~km~s$^{-1}$, a gas density of $10$~cm$^{-3}$, an electron temperature of $8000$~K and an emission measure of $1000$~pc~cm$^{-6}$, equivalent to $L=10$~pc and $f_{\mathrm{m}}=5.5\times10^{-4}$. For this validation exercise we adopt a diameter of $18$~kpc for the Galactic disk (in this exercise the exact value is not important, as it is equivalent to varying $f_{\mathrm{m}}$ and $f_{\mathrm{bg}}$), and a distance of $10$~kpc for the gas, so we have $f_{\mathrm{bg}}=0.44$. For the continuum we adopt a powerlaw with $T_{0}=32.9$~K at 1~GHz and $\beta=-2.2$. We add random noise drawn from a Normal distribution to the synthetic spectra to reproduce the mean S/N of our observations.

To explore the correlations between parameters we sample their distributions using uniform priors in logarithmic space, except for $f_{\mathrm{bg}}$ which is sampled in linear space. We explore the cases where $f_{\mathrm{bg}}$ and $f_{\mathrm{m}}$ are unknown by setting them to take values between $0$ and $1$ and $10^{-2}$ and $10^{-4}$, respectively. We also explore the cases with $f_{\mathrm{m}}=L/D_{\rm{gal}}$ and $f_{\rm{bg}}=(D_{\rm{gal}}-D)/D_{\rm{gal}}$.

The results of this analysis are presented in Figure~\ref{fig:synth_corner}. We find strong correlations between $T_{\rm{e}}$, EM and $v_{\rm{rms}}$. The correlation between $T_{\rm{e}}$ and EM arises from the dependence of the line brightness on these quantities. As for a higher temperature a larger EM is required to explain the line brightness. The correlation between $T_{\rm{e}}$ and $v_{\rm{rms}}$ is due to the line width. We also find significant correlations between $f_{\rm{bg}}$ and the rest of the parameters, except for $f_{\mathrm{m}}$. The strongest of these is with $n_{\mathrm{e}}$, which is due to the effect of stimulated emission. For a stronger background continuum, larger $f_{\rm{bg}}$, less stimulated emission, higher $n_{\mathrm{e}}$, is required to explain the line brightness. $f_{\mathrm{m}}$ does not show a correlation with the remaining parameters, and its effect is to shift the maximum of the posteriors. For all the parameterizations of $f_{\mathrm{bg}}$ and $f_{\mathrm{m}}$ the maximum in the posterior distribution for $n_{\mathrm{e}}$ is biased towards higher values by $\approx20\%$, with the input value inside the $2\sigma$ contour of the posterior.

Figure~\ref{fig:synth_corner} also shows that the posterior distributions for $n_{\mathrm{e}}$ have well defined peaks and that these peaks are within $2\sigma$ of the input value, for all the parametrizations of $f_{\mathrm{m}}$, $f_{\mathrm{bg}}$ and $T_{\rm{e}}$. For this reason, we consider it to be the only parameter constrained by the model and observations. Thus, we also explore the posterior distributions if we adopt an informative prior on the gas temperature. We use a normal distribution for $T_{\mathrm{e}}$ centered on the input value and with a width of $350$~K, taken from the Galactic temperature curve for a known distance \citep{Quireza2006}. In this case the correlation between parameters are no longer significant and the bias in $n_{\mathrm{e}}$ is reduced to $6\%$.

In summary, this validation experiment shows that modeling HRRLs at $342$, $800$ and $5757$~MHz constrains $n_{\mathrm{e}}$, with a bias of $\approx20\%$, and does not constrain the $T_{\mathrm{e}}$, EM, $v_{\mathrm{rms}}$, $T_{\mathrm{bg}}$ or $T_{\mathrm{nth}}$. External constraints or assumptions are required to constrain these parameters, which reduces the bias in $n_{\mathrm{e}}$.

\subsection{Gas properties}
\label{ssec:tb}

We apply our method to the observations of the three regions. We adopt a normal prior on the gas temperature, with a mean and width given by the Galactic temperature curve of \citet{Quireza2006} evaluated at the distances in Table~\ref{tab:kdist}. We parametrize $f_{\mathrm{bg}}=(D_{\mathrm{gal}}-D)/D_{\mathrm{gal}}$ and $f_{\mathrm{m}}=L/D_{\mathrm{gal}}$.

\begin{deluxetable*}{lcccccc}
\digitalasset
\tablewidth{0pt}
\tablecaption{Derived gas properties \label{tab:gasprops}}
\tablehead{
\colhead{Region} & \colhead{ID} & \colhead{$n_{\mathrm{e}}$ (cm$^{-3}$)} & \colhead{EM (pc cm$^{-6}$)} & \colhead{T (K)} & \colhead{$v_{\mathrm{rms}}$ (km s$^{-1}$)} & \colhead{L (pc)}
}
\startdata
\multicolumn{7}{c}{Near KD} \\
\hline
G15.85+0.10 & 1 & $11.3_{-0.4}^{+0.4}$ & $3005_{-249}^{+268}$ & $5815_{-328}^{+343}$ & $17.0_{-0.3}^{+0.3}$ & $23_{-2}^{+2}$ \\
        \multirow{2}{*}{G24.07$-$0.59} & 1 & $9.7_{-0.6}^{+0.6}$ & $1541_{-129}^{+143}$ & $6741_{-374}^{+399}$ & $11.7_{-0.4}^{+0.4}$ & $16_{-2}^{+3}$ \\
                                     & 2 & $11.6_{-0.5}^{+0.5}$ & $1769_{-142}^{+160}$ & $5164_{-280}^{+308}$ & $17.0_{-0.3}^{+0.3}$ & $13_{-1}^{+2}$ \\
        G31.00$-$0.91 & 1 & $14.4_{-0.7}^{+0.7}$ & $2899_{-232}^{+257}$ & $6547_{-360}^{+382}$ & $21.2_{-0.3}^{+0.3}$ & $14_{-2}^{+2}$ \\ 
		\hline
        \multicolumn{7}{c}{Far KD} \\
        \hline
        G15.85+0.10 & 1 & $7.6_{-0.4}^{+0.2}$ & $3160_{-260}^{+280}$ & $5850_{-380}^{+300}$ & $17.4_{-0.3}^{+0.3}$ & $56_{-7}^{+5}$ \\
        \multirow{2}{*}{G24.07$-$0.59} & 1 & $6.7_{-0.4}^{+0.4}$ & $1578_{-132}^{+161}$ & $6623_{-364}^{+431}$ & $12.0_{-0.4}^{+0.4}$ & $36_{-5}^{+6}$ \\
                                     & 2 & $9.7_{-0.4}^{+0.4}$ & $1792_{-143}^{+157}$ & $5147_{-278}^{+301}$ & $17.1_{-0.3}^{+0.3}$ & $19_{-2}^{+2}$ \\
        G31.00$-$0.91 & 1 & $12.4_{-0.7}^{+0.7}$ & $2934_{-237}^{+260}$ & $6517_{-358}^{+384}$ & $21.3_{-0.3}^{+0.3}$ & $19_{-2}^{+3}$ \\ 
\enddata
\end{deluxetable*}

\begin{figure*}
    \includegraphics[width=\linewidth]{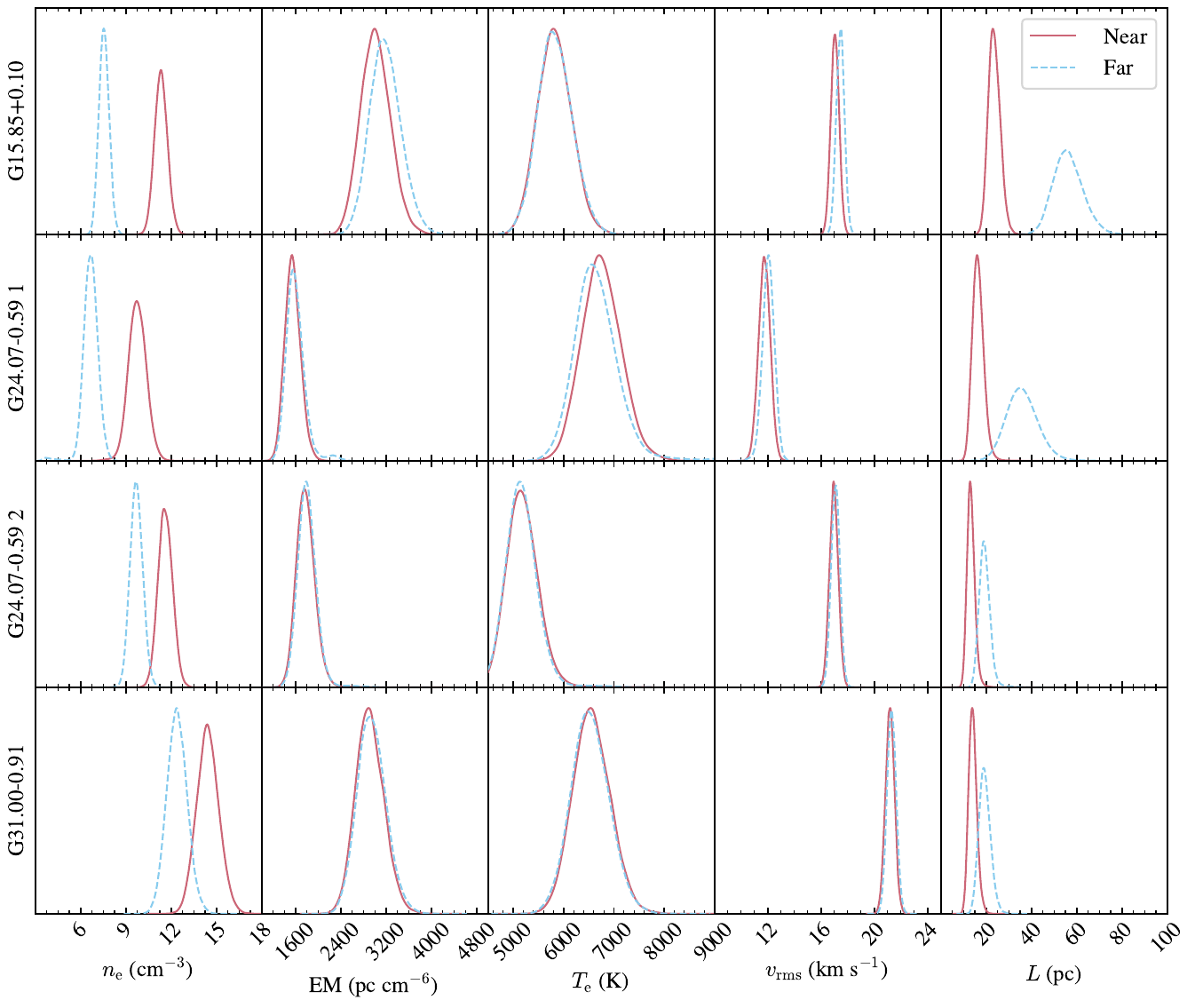}
    \caption{Posterior distributions for the gas properties at the near and far KD. Each row shows the posterior distributions for both KD solutions for one velocity component. From top to bottom; G15.85+0.10, G24.07$-$0.59 component 1, G24.07$-$0.59 component 2, and G31.00$-$0.91.}
    \label{fig:gasprops}
\end{figure*}

The derived gas properties are presented in Table~\ref{tab:gasprops} and Figure~\ref{fig:gasprops}, where we include values for both solutions to the kinematic distance ambiguity. As shown in Sect.~\ref{ssec:validation}, placing the gas at the near kinematic distance has the effect of raising the electron density. The range of electron density between the two KD solutions gives a sense for the uncertainty for the electron density if the adopted distances are in error. The most extreme case is G15.85+0.10, for which a change in KD from $3$ to $13$~kpc results in a change in $n_{\mathrm{e}}$ of $33\%$.

In Table~\ref{tab:gasprops} we also include gas properties for the $90$~km~s$^{-1}$ component towards G24.07$-$0.59, however, these should be interpreted with caution as we have not accounted for the effects of beam dilution. As we will discuss in Section~\ref{ssec:uncertainties}, if we were to correct the data for beam dilution, then the derived densities would be lower.

The derived electron densities are between $6$ and $15$~cm$^{-3}$, the emission measures between $1500$ and $3200$~cm$^{-3}$, which results in path lengths between $13$ and $56$~pc. For these densities, and given the priors on the temperature, the gas thermal pressure ($P=n_{\mathrm{e}}T_{\mathrm{e}}$) is of the order of $6\times10^{5}$~K~cm$^{-3}$. This pressure is closer to the thermal pressures found for the ionized gas in \Hii\ regions \citep[e.g.,][]{Pabst2020} than to that of the WIM \citep[$\sim800$~K~cm$^{-3}$,][]{Haffner2009}. The recombination timescale for gas at a density of $n_{\mathrm{e}}\approx10$~cm$^{-3}$ is $\approx10^{4}$~yr.

\subsection{Uncertainties}
\label{ssec:uncertainties}

Throughout our analysis of the line intensities we have assumed that the gas fills the beam, and that the HRRL emission comes from a single homogeneous gas volume. Here we explore how deviations from these assumptions affect the derived gas properties in a general case. That is, we do not aim to quantify how our derived gas properties would change if these assumptions are not valid for the lines of sight studied, but to illustrate how deviations from these assumptions affect the derived gas properties.

To explore the effect of the beam filling factor we use the synthetic observations presented in Section~\ref{ssec:validation}, and assume that the emission comes from a smaller solid angle. This is equivalent to increasing the line intensity. When the emission comes from a smaller area than that of the $800$~MHz observations, EM increases, with no changes in $n_{\mathrm{e}}$. If the area of the HRRL emission is larger than the area of the $800$~MHz observations but smaller than that of the $342$~MHz observations, then $n_{\mathrm{e}}$ decreases.

We explore the effect of having different volumes of gas responsible for the HRRL emission. For this, we use a simple model in which there are two volumes of gas within the beam. The only differences between the volumes are their electron densities and EMs. We add the brightness of the HRRLs coming from the two volumes. We start with the case where the EMs remain the same. We consider a volume with $n_{\mathrm{e}}=10$~cm$^{-3}$ and the other with $n_{\mathrm{e}}=15$~cm$^{-3}$. In this case, we derive an electron density of $n_{\mathrm{e}}=17$~cm$^{-3}$. If the densities of the gas are significantly different, but with similar EM then we derive $n_{\rm{e}}$ close to the mean of the densities of the two volumes. For example, for $n_{\mathrm{e}}=10$~cm$^{-3}$ and $100$~cm$^{-3}$ with EM$=1000$~pc~cm$^{-6}$ we derive $n_{\mathrm{e}}=61$~cm$^{-3}$. If in the previous example we increase the EM for the high density component by a factor of five, then it is not possible to find convergence for the electron density. All in all, this shows that solutions will only be possible when the EMs are comparable, and in that case the derived $n_{\mathrm{e}}$ will be close to the EM weighted average $n_{\mathrm{e}}$ of the volumes. In all these examples we have assumed that the continuum radiation from one of the volumes responsible for the HRRL emission does not contribute to the total continuum from the other volume -- the volumes do not see each other. The results are similar if we average the line intensities.

We also investigate the effect of our choice of departure coefficients on the derived gas properties by using the line properties reported by \citet{Roshi2001} and their methods used to derive $n_{\mathrm{e}}$. As an example, we use the region centered on $\ell=16.1\arcdeg$ -- the closest to G15.85+0.10. For this region \citet{Roshi2001} report $n_{\mathrm{e}}=6.1$~cm$^{-3}$ and $2.9$~cm$^{-3}$ for distances of $2.8$~kpc and $13.5$~kpc, respectively. If we take their data for this region (Tables~1 and 2 in \citealt{Roshi2001}), their beam filling factors, use the same procedure as described in their Sect.~3.1, and the departure coefficients of \citet{Salgado2017a} to derive the gas properties we find $n_{\mathrm{e}}\approx2$~cm$^{-3}$ at a distance of $13.5$~kpc, which is consistent with their result ($2.9$~cm$^{-3}$). The $30\%$ difference can be attributed to the choice of departure coefficients, and indicates that for the same inputs using the departure coefficients of \citet{Salgado2017a} results in lower values of $n_{\mathrm{e}}$.

\subsection{Measuring EM and $T_{\mathrm{e}}$}
\label{ssec:degeneracy}

The emission measure and temperature of the ionized gas probed by HRRLs outside of \Hii\ regions hold important clues about the ISM. Although there is consensus that broad HRRLs (line widths $\gtrsim15$~km~s$^{-1}$) probe warm ($T_{\mathrm{e}}\sim5000~K$) gas \citep[e.g.,][]{Shaver1976}, being able to accurately determine EM and $T_{\mathrm{e}}$ from HRRL observations would improve our ability to characterize the warm ionized gas. Our validation exercise (Sect.~\ref{ssec:validation}) shows that without an informative prior on the gas temperature, or emission measure, the gas EM and $T_{\mathrm{e}}$ cannot be accurately measured.

To illustrate the importance of having an accurate measurement of the gas temperature we consider the temperature measurements of \citet{Langer2021}. Through observations of HRRLs and the FIR \Nii\ line they derive temperatures in the range $3.4\times10^{3}$ to $8.5\times10^{3}$~K for $20.8\arcdeg<\ell<28.7\arcdeg$ and $b=0\arcdeg$ by comparing the line widths of both lines. This is more than a factor of two variation in temperature over a fraction of the Galactic disk (their measurements lie between $R_{\mathrm{gal}}$ $4$ and $5$~kpc), and outside the ranges expected from measurements of the temperature towards \Hii\ regions at similar locations in the Galaxy \citep{Balser2015}. A variation of a factor of $2$ in temperature results in a $\approx2.8$ variation in EM, and path length (Fig.~\ref{fig:synth_corner}).

\begin{figure}
    \centering
    \includegraphics[width=\linewidth]{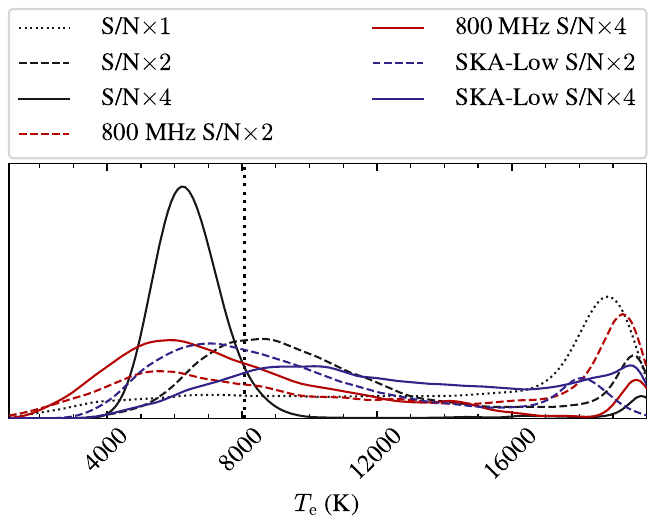}
    \caption{Effects of increasing the S/N of one or all measurements in the posterior distribution for $T_{\mathrm{e}}$. The dotted line shows the base case, with a S/N comparable to that of the observations presented in this work. The black dashed and solid lines show the effect of increasing the S/N of all measurements by factors of two and four, respectively. The red dashed and solid lines show the effect fo increasing the S/N of the measurements at $800$~MHz by factors of two and four, respectively. The blue dashed and solid lines show the effect of including a measurement at $182$~MHz with a S/N factors of two and four larger than that of the observations presented here, respectively.}
    \label{fig:synthsn}
\end{figure}

We explore the possibility of improving the measurement of EM and $T_{\mathrm{e}}$ through observations of HRRLs, like those presented here, through increasing the S/N of the observations or adding additional lines. We use the same setup as that used during the method validation, with no assumptions on the relation between $f_{\mathrm{m}}$ and $f_{\mathrm{bg}}$ on $L$ or the distance. To quantify our results we focus on the posterior distributions for EM and $T_{\mathrm{e}}$. In particular how the $1\sigma$ ranges change, and what is the difference between the mean and input values. The resulting posterior distributions for $T_{\mathrm{e}}$ from this analysis are presented in Fig.~\ref{fig:synthsn}.

If the S/N for all HRRLs was increased by a factor of two, then there would be a marginal reduction in the $1\sigma$ uncertainties, but the peaks of the posteriors for EM and $T_{\mathrm{e}}$ would be closer to their input values. The bias in EM would be reduced from $160\%$ to $30\%$, and from $100\%$ to $20\%$ for $T_{\mathrm{e}}$. The $1\sigma$ uncertainties would be reduced from $126\%$ to $100\%$ and from $70\%$ to $60\%$ for EM and $T_{\mathrm{e}}$, respectively. For a factor of four increase in the S/N, the $1\sigma$ uncertainties would be reduced to $\approx15\%$ and the bias to $25\%$.

If we consider only increasing the S/N for one of the HRRLs, then we have that if we doubled the S/N of the $800$~MHz observations, then the bias would reduce to $55\%$ and $40\%$, and the uncertainties to $140\%$ and $85\%$ for EM and $T_{\mathrm{e}}$, respectively. Increasing the S/N by a factor of four at $800$~MHz would reduce the bias to $15\%$ and the uncertainty to $60\%$. Increasing the S/N for the $342$~MHz or $5757$~MHz observations would result in lower reductions in the biases and uncertainties, although the specific reductions are dependent on the gas density.

We also explore the effect of adding an additional HRRL in the frequency range expected to be covered by SKA-Low ($50$ to $350$~MHz). We arbitrarily pick a HRRL at $182$~MHz with a principal quantum number of $330$. Including this HRRL with a S/N of $240$ (four times the S/N of the observations presented here) would reduce the bias to $60\%$ and the uncertainties would be reduced by $50\%$.

Thus, to independently constrain the temperature of the ionized gas at densities around $10$~cm$^{-3}$, the most efficient way seems to be to increase the S/N of observations around $800$~MHz. For regions like G15.85+0.10 this would imply spending three hours on source, for fainter regions, like G31.00$-$0.91, this would require roughly ten hours of on-source time. Another potential advantage of increasing the S/N would be the potential detection of helium RRLs. By comparing the line widths of helium and hydrogen RRLs it is possible to measure the gas temperature. Although high S/N detections of helium RRLs are needed for this approach to yield precise temperatures.

\subsection{Comparison to previous HRRL observations}
\label{ssec:previouswork}

Our analysis is based on the methods developed by \citet{Shaver1975} and then applied to low-frequency HRRLs by \citet{Anantharamaiah1985b} and \citet{Roshi2001}. Given the similarity of the methods we investigate the differences and similarities between our results.

\citet{Anantharamaiah1985b} used the H$272\alpha$ line at $324.99$~MHz and the H$166\alpha$ line at $1424.73$~MHz \citep{Lockman1976} to study the ionized gas towards regions of faint continuum emission. To identify regions of faint continuum emission, blank regions, they used the $5$~GHz maps of \citet{Altenhoff1979}. The blank regions are located at $|b|\leq0.3\arcdeg$ and $2.1\arcdeg\leq\ell\leq21.2\arcdeg$, and the beam of the $324.99$~MHz observations was $2\arcdeg\times6\arcmin$. From their analysis they derive a mean electron density of $2.8$~cm$^{-3}$ with a $1\sigma$ scatter of $1.9$~cm$^{-3}$. \citet{Roshi2001} used four HRRLs around $327$~MHz along $1.4$~GHz HRRLs to study the ionized gas in the Galactic plane ($b\sim0\arcdeg$) over the Galactic latitude range $\ell=330\arcdeg$ to $89\arcdeg$ at an angular resolution of $\approx2\arcdeg\times2\arcdeg$. They derive electron densities between $1$~cm$^{-3}$ and $16$~cm$^{-3}$, a larger range than that presented by \citet{Anantharamaiah1985b}. The density derived towards G15.85+0.10, for the near KD, is higher than those reported by \citet{Roshi2001} towards similar Galactic longitudes. For G31.00$-$0.91 we derive an electron density which is larger than those found by \citet{Anantharamaiah1985b}, and at the high end of those reported by \citet{Roshi2001}, regardless of the adopted KD. A more detailed comparison requires maps of the regions analyzed by \citet{Anantharamaiah1985b} and \citet{Roshi2001}.

\subsection{Comparison with FIR observations}

The ionized gas in the Galactic plane has also been studied using far-infrared (FIR) observations of ionized nitrogen \citep[e.g.,][]{Goldsmith2015,Langer2021}. Using the PACS instrument on-board Herschel \citet{Goldsmith2015} derived electron densities using the ratio of the fine structure lines of \Nii. They found $n_{\mathrm{e}}$ in the range $10$-$50$~cm$^{-3}$. For their pointing towards G015.7+0.0 they report $n_{\mathrm{e}}=18.6\pm2.1$~cm$^{-3}$, and for G031.3+0.0 $31.2\pm0.5$~cm$^{-3}$. These values are a factor of $1.5$ to two larger than the values we derive. For the column density, we find values which are a factor of $3$ to $9$ larger than those derived from Table~2 of \citet{Goldsmith2015} adopting their representative nitrogen abundance of $2.9\times10^{-4}$.

Another method used to characterize ionized gas leverages the ratio between HRRLs and the FIR \Nii\ lines, the RRL–FS method \citep[e.g.,][]{Pineda2019}. For G30.0+0.0 \citet{Pineda2019} report $n_{\mathrm{e}}=17.6\pm9.9$~cm$^{-3}$, which is consistent with the value we find for G31.00$-$0.91. 

These differences suggest that the HRRLs used in this analysis probe gas to lower densities than those probed through the FIR \Nii\ lines. This would also explain the higher column densities and path lengths implied by HRRL observations, which could be due to the presence of additional gas at low density not probed by \Nii. A larger sample would help establish if low-frequency HRRLs are probing lower density gas relative to that probed by \Nii.

\subsection{Ionization}

Given the values of $n_{\rm{e}}$ and $T_{\mathrm{e}}$ we can estimate the number of Lyman continuum photons, $N_{\rm{Lyc}}$, required to photoionize the gas probed by HRRLs. We follow \citet{Emig2020b} to estimate $N_{\rm{Lyc}}$ from the electron density and temperature of the gas (Table~\ref{tab:gasprops}). For these calculations we adopt the solutions derived assuming the near kinematic distances. If we adopt a volume filling factor of unity, and assume that the gas occupies a cylindrical volume with a base of $16\arcmin$ and a height given by $L$ (Table~\ref{tab:gasprops}), then we have $N_{\rm{Lyc}}=(5.6\pm0.6)\times10^{48}$~s$^{-1}$ and $(1.3\pm0.2)\times10^{49}$~s$^{-1}$ for G15.85+0.10 and G31.00$-$0.91, respectively. For G24.07$-$0.59 we have $(3.3\pm0.6)\times10^{48}$~s$^{-1}$ and $(1.2\pm0.2)\times10^{49}$~s$^{-1}$ for the $54$~km~s$^{-1}$ and $97$~km~s$^{-1}$ components, respectively. We compare these ionization rates to those measured for \Hii\ regions close to the targets. We use the catalog of \citet{Murray2010}, which provides distances and luminosities for \Hii\ regions. 

Close to G15.85+0.10 the catalog lists a source at $(\ell,b)=(14.7\arcdeg,-0.5\arcdeg)$ and a distance of $3.5$~kpc. This source has a luminosity of $2\times10^{24}$~erg~s$^{-1}$~Hz$^{-1}$, which corresponds to a ionization rate of $2.7\times10^{50}$~s$^{-1}$. Considering the distance between the \Hii\ region and G15.85+0.10, this implies that the contribution from this \Hii\ region to the ionization of G15.85+0.10 is $1.8\times10^{46}$~s$^{-1}$. This is likely an upper limit, as we do not consider any absorption between the source and target. Thus, this region alone could not be the sole source of ionizing photons responsible for the ionization of G15.85+0.10. Even if we assume the same distance ($3.07$~kpc) for both, the contribution would be $7\times10^{47}$~s$^{-1}$. M17 and M16 are at a larger distance from G15.85+0.10, thus their contributions to the ionization are lower. 

For G31.00$-$0.91 we consider the contribution from W43. Based on the results of \citet{Murray2010} this \Hii\ region produces $3.1\times10^{51}$~s$^{-1}$ ionizing photons, or $2.5\times10^{51}$~s$^{-1}$ if we adopt the parallax distance to W43 \citep{Zhang2014}. This results in a contribution of $3.6\times10^{47}$~s$^{-1}$ ionizing photons, using the parallax distance, to the ionization of G31.00$-$0.91. If we assume that W43 and G31.00$-$0.91 are at the same distance ($5.02$~kpc), then the contribution to the ionizing photons is $8.4\times10^{48}$~s$^{-1}$, which is consistent with the ionization rate for G31.00$-$0.91.

Towards $\ell=24.5\arcdeg$ \citet{Murray2010} list $14$ \Hii\ regions, of which $8$ are at distances between $2$ and $7$~kpc. These produce a combined photon ionization rate of $3.3\times10^{51}$~s$^{-1}$, and contribute $1.6\times10^{46}$~s$^{-1}$ and $6.6\times10^{47}$~s$^{-1}$ to the ionization of the $54$~km~s$^{-1}$ and $97$~km~s$^{-1}$ components, respectively. These contributions are a factor of $\sim100$ lower than the derived ionization requirements for the velocity components observed towards G24.07$-$0.59.

In summary, for three out of the four velocity components analyzed the ionization requirements cannot be explained by the \Hii\ regions listed by \citet{Murray2010}.

\section{Summary}
\label{sec:summary}

We presented new observations of low-frequency HRRLs at $342$ and $800$~MHz towards three positions in the Galactic plane, two of which are devoid of known \Hii\ regions. We detect HRRLs in all positions at velocities similar to those observed in the $^{12}$CO$(1\mbox{-}0)$ line, suggesting that the HRRLs are probing gas associated with the spiral arms of the Galaxy.

We combine our HRRL observations with higher frequency ($\nu\approx5.8$~GHz) transitions from GDIGS and ancillary data to determine the properties of the ionized gas extending the methods developed by \citet{Shaver1975} and \citet{Anantharamaiah1985b} to model the line profiles from input gas properties. Our analysis shows that using HRRLs in this frequency range the electron density can be determined to within $\approx20\%$, and that its value is only weakly dependent on other parameters. The remaining properties (i.e., temperature and emission measure) can be determined with the aid of additional constraints, or through higher sensitivity measurements of HRRLs over a similar frequency range.

We find that towards two of the three positions observed the HRRLs can be explained by a homogeneous volume of gas filling the beam with $n_{\mathrm{e}}$ in the range $6$ to $15$~cm$^{-3}$ and emission measure between $1500$ and $3200$~pc~cm$^{-6}$ for gas with a temperature between $5000$ and $6800$~K. For the third position, the differences in angular resolution complicate the analysis, highlighting the need for observations at the same angular resolution or sampling the same areas to perform multi-frequency comparisons. 

The range of $n_{\mathrm{e}}$ is higher than that derived from previous low-frequency HRRL observations \citep[e.g.,][]{Anantharamaiah1985b,Roshi2001}, and lower than those found through FIR observations of \Nii\ \citep[e.g.,][]{Goldsmith2015,Langer2021} and the RRL–FS method \citep[e.g.,][]{Pineda2019}. The difference with respect to previous HRRL determinations can be explained by the unknown beam filling factors of previous HRRL studies, and shows the benefits of using fully spatially sampled HRRL cubes to estimate the beam filling factor. The difference with respect to \Nii\ could be explained by additional low density material along the lines of sight that is not probed by \Nii. Future observations with the GBT, and other sensitive low-frequency facilities (e.g., NenuFAR, LOFAR, SKA, FAST), will enable an accurate characterization of the ionized gas density in the Galactic plane over large scales.

\begin{acknowledgments}
The authors would like to thank the referee for constructive and positive feedback. 
The National Radio Astronomy Observatory and Green Bank Observatory are facilities of the U.S. National Science Foundation operated under cooperative agreement by Associated Universities, Inc.
PS \& KLE would like to welcome Liam Emig Salas into this world.
This research made use of Astropy, a community-developed core Python package for Astronomy \citep{Astropy2013,Astropy2018}, matplotlib, a Python library for publication quality graphics \citep{Hunter2007}, LMFIT, a nonlinear least-square minimization and curve-Fitting package for Python \citep{newville_2014_11813}, CRRLpy, a Python package for the analysis of RRL observations \citep{crrlpy}, and \texttt{ChainConsumer} \citep{Hinton2016}. The research presented in this paper has used data from the Canadian Galactic Plane Survey, a Canadian project with international partners, supported by the Natural Sciences and Engineering Research Council.
\end{acknowledgments}

\vspace{5mm}
\facilities{Green Bank Telescope (PF800, PF342, C).}

\software{Astropy \citep{Astropy2013,Astropy2018},  
          CRRLpy \citep{crrlpy},
          Matplotlib \citep{Hunter2007},
          LMFIT \citep{newville_2014_11813},
          ChainConsumer \citep{Hinton2016},
          dysh \citep{Pound2025}
          }

\bibliographystyle{aasjournal}
\bibliography{refs.bib}{}

\end{document}